\documentclass[aps,amssymb,amsmath,twocolumn,pra,superscriptaddress]{revtex4}
\usepackage[pdftex]{graphicx}
\usepackage{dcolumn}
\usepackage{bm}
\usepackage{bbm}
\usepackage{hyperref}
\usepackage{color}
\usepackage[up]{subfigure}

\newcommand{\be}{\begin{equation}}
\newcommand{\ee}{\end{equation}}
\newcommand{\bc}{\begin{center}}
\newcommand{\ec}{\end{center}}
\newcommand{\bea}{\begin{eqnarray}}
\newcommand{\eea}{\end{eqnarray}}
\newcommand{\ba}{\begin{array}}
\newcommand{\ea}{\end{array}}

\begin{document}

\title{Decoherence in two-dimensional quantum walks using four- and
  two-state particles}

\author{C. M. Chandrashekar}
\email{c.madaiah@oist.jp}

\affiliation{Department of Physics, University College Cork, Cork,
  Republic of Ireland}

\affiliation{Quantum Systems Unit, Okinawa Institute of Science and
  Technology Graduate University, Okinawa, Japan}

\author{Th. Busch}
\email{thomas.busch@oist.jp}

\affiliation{Department of Physics, University College Cork, Cork,
  Republic of Ireland}

\affiliation{Quantum Systems Unit, Okinawa Institute of Science and
  Technology Graduate University, Okinawa, Japan}


\begin{abstract}
  We study the decoherence effects originating from state flipping and depolarization for two-dimensional discrete-time quantum walks using four-state and two-state particles. By quantifying the quantum correlations between the particle and position degree of freedom and between the two spatial ($x-y$) degrees of freedom using measurement induced disturbance (MID), we show that the two schemes using a two-state particle are more robust against decoherence than the Grover walk, which uses a four-state particle. We also show that the symmetries which hold for two-state quantum walks breakdown for the Grover walk, adding to the various other advantages of using two-state particles over four-state particles.
\end{abstract}
\maketitle

\section{Introduction}
\label{intro}

Quantum walks are a close quantum analog of classical random walks, in
which the evolution of a particle is given by a series of superpositions in position 
space\,\cite{Ria58, Fey86, Par88, LP88, ADZ93}. Recently they have emerged as 
an efficient tool to carry out quantum algorithms\,\cite{Kem03,Amb03} and have
been suggested as an explanation for wavelike energy transfer within
photosynthetic systems\,\cite{ECR07, MRL08}.  They have applications
in the coherent control of atoms and Bose-Einstein condensates in
optical lattices\,\cite{CL08, Cha11a}, the creation of topological phases\,\cite{KRB10}, and the generation of entanglement\,\cite{GC10}. Quantum walks therefore have the potential to serve as a
framework to simulate, control and understand the dynamics of a variety of physical and biological systems. Experimental implementations of quantum walks in last few years have included
NMR\,\cite{DLX03, RLB05, LZZ10}, cold ions\,\cite{SMS09, ZKG10},
photons\,\cite{PLP08, PLM10, SCP10, BFL10, SCP11, SSV12}, and
ultracold atoms\,\cite{KFC09}, which has drawn further interest of the
wider scientific community to their study.

The two most commonly studied forms of quantum walks are the
continuous-time\,\cite{FG98} and the discrete-time
evolutions\,\cite{ADZ93, DM96, ABN01, NV01, Kon02, BCG04, CSL08}. In
this work we will focus on the discrete-time quantum walk and just
call it quantum walk for simplicity. If we consider a one-dimensional (1D)
example of a two-state particle initially in the state
\begin{equation}
  |\Psi_{\rm in}\rangle= \left( \cos(\delta/2)| 0 \rangle +
                       e^{i\eta}\sin(\delta/2)| 1 \rangle \right)
                       \otimes |\psi_{0}\rangle,
\end{equation}
then the operators to implement the walk are defined on the
{\it coin} (particle) Hilbert space $\mathcal{H}_c$ and the {\it
  position} Hilbert space $\mathcal{H}_p$ [$\mathcal{H} =
\mathcal{H}_c \otimes \mathcal{H}_p$]. A full step is given by first using the
the unitary quantum coin 
\begin{equation}
\hat{B} (\theta) \equiv 
        \begin{bmatrix}\cos(\theta)   &   ~~\sin(\theta)  \\
                       \sin(\theta)   & -\cos(\theta)  
        \end{bmatrix},
\end{equation}
and then following it by a conditional shift operation 
\begin{equation}
  \hat{S}_x \equiv   \sum_x \left [  |0 \rangle\langle
    0|\otimes|\psi_{x-1}\rangle\langle   \psi_x|   +  |1 \rangle\langle
    1|\otimes |\psi_{x+1}\rangle\langle \psi_x| \right ].
\end{equation}
The state after $t$ steps of evolution is therefore given by
\begin{equation}
   |\Psi_t\rangle= [\hat{S}_x [\hat{B}(\theta) \otimes 
                    \hat{{\mathbbm 1}}]]^t|\Psi_{\rm in}\rangle\;.  
\end{equation}
All experimental implementations of quantum walks reported by today
have used effectively 1D dynamics.  A natural extension
of 1D quantum walks to higher dimension is to enlarge the Hilbert
space of the particle with one basis state for each possible direction
of evolution at the vertices. Therefore, the evolution has to be
defined using an enlarged coin operation followed by an enlarged
conditioned shift operation. For a two-dimensional (2D) rectangular
lattice the dimension of the Hilbert space of the particle will be
four and a four dimensional coin operation has to be used.  Two
examples of this are given by using either the degree four discrete
Fourier operator (DFO) [Fourier walk] or the Grover diffusion operator
(GDO) [Grover walk] as coin operations\,\cite{MBS02,
  TFM03, HGJ11}. An alternative extension to two and higher $(d)$ dimensions
is to use $d$ coupled qubits as internal states to evolve the
walk\,\cite{EMB05, OPD06}. Both these methods are experimentally demanding
 and beyond the capability of current experimental set ups. Surprisingly however, two 
alternative schemes to implement quantum walks on a 2D lattice were recently proposed which
use only two-state particles.  In one of these a single two-state
particle is evolved in one dimension followed by the evolution in
other dimension using a Hadamard coin operation\,\cite{FGB11, FGM11}. In
the other, a two-state particle is evolved in one dimension followed
by the evolution in the other using basis states of different Pauli
operators as translational states\,\cite{CBS10, Cha12}.

In this work we expand the understanding of 2D quantum walks by studying the effects decoherence has on the four-state Grover walk and the two two-state
walks mentioned above. The environmental effects are modeled using a
state-flip and a depolarizing channel and we quantify the quantum
  correlations using a measure based on the disturbance induced by
  local measurements \cite{L08}.  While in the absence of noise
    the probability distributions for all three schemes are
    identical, the quantum correlations built up during the evolutions
    differ significantly. However, due to the difference in the size of the particles 
Hilbert space for the Grover walk and the two-state walks, quantum correlations 
generated between the particle and the position space cannot be compared. The quantum correlations between the two spatial dimensions ($x-y$), obtained
  after tracing out the particle state, on the other hand, can be compared and we will show that they are larger for 
the walks using the two-state particles. When taking the environmental effects into account, we
  find that all three schemes lead to different probability
  distribution and decoherence is strongest for the Grover walk, therefore making the two-state walks more
    robust for maintaining quantum correlations.
Interestingly, we also find that certain symmetries which hold for the two-state quantum walk in the presence of noise do
not hold for a Grover walk. Together with the specific
initial state and the coin operation required for the evolution of the
Grover walk, this reduces the chances to identify an equivalence class of
operations on a four-state particle to help experimentally implement the
quantum walk in any physical system that allows to
manipulate the four internal states of the coin.

This article is organized as follows : In Section\,\ref{2dqw} we
define the three schemes for the 2D quantum walk used to study the
decoherence and in Section\,\ref{qmid}
  we define the measure we use to quantify the quantum correlations. In Section\,\ref{decoqw}, the effect
of decoherence in the presence of a state-flip noise channel and a depolarizing
channel are presented and we compare the quantum correlations 
between the $x$ and $y$ directions for the three schemes. We finally show in Section\,
\ref{sec:symm} that the state-flip and phase-flip symmetries, which
hold for the two-state quantum walk, breakdown for the four-state walk
and conclude in Section\,\ref{conc}.

\section{Two-dimensional quantum walks}
\label{2dqw}

\subsection{Grover walk}
\label{4sqw}

\begin{figure}
  \includegraphics[width=8.0cm]{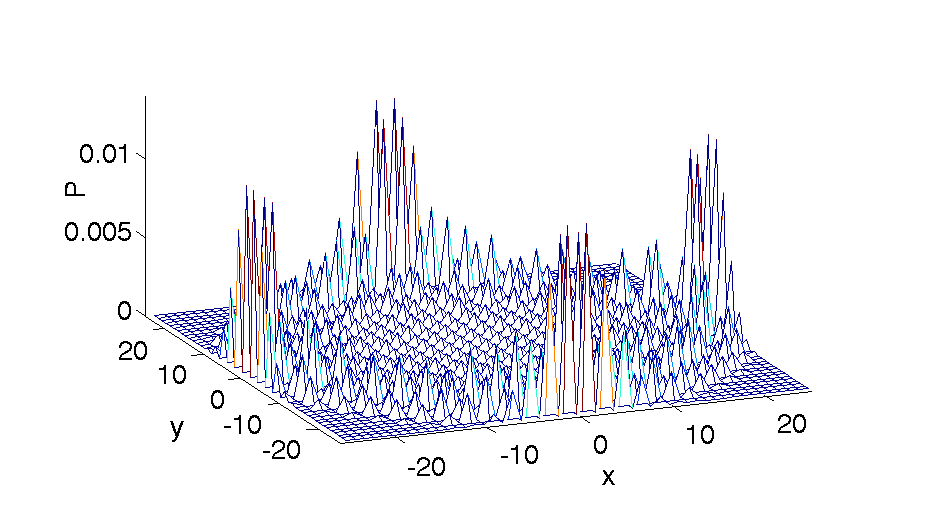}
   \caption{(Color online) Probability distribution for the Grover after 25 steps. 
    An identical distribution using the two-state particle with the  $\frac{|0\rangle + i |1\rangle}{\sqrt 2}$ can be obtained using the alternate walk with the coin operation $B(\pi/4)$ and the Pauli walk after 25 steps.}
    \label{fig:1}
\end{figure}

For a Grover walk of degree four the coin operation is given
by\,\cite{MBS02, TFM03}
\begin{equation}
\hat{G} = \frac{1}{2}\begin{bmatrix}-1 & ~~1 & ~~1 & ~~1  \\
  ~~1 & -1 & ~~1 & ~~1  \\
  ~~1 & ~~1 & -1 & ~~1  \\
  ~~1 & ~~1 & ~~1 & -1
\end{bmatrix},
\label{grovercoin}  
\end{equation}
and the shift operator is 
\begin{align}
  \hat{S}_{(x, y)} \equiv& \sum_{x, y} \Big[|0 \rangle\langle 0|\otimes|
                          \psi_{x-1, y-1}\rangle\langle \psi_{x, y}| \nonumber \\  
                        &\;\;\;+     |1 \rangle\langle 1|\otimes |
                          \psi_{x-1, y+1}\rangle\langle \psi_{x, y}| \nonumber\\
                        &\;\;\;+     |2 \rangle\langle 2|\otimes|
                          \psi_{x+1, y-1}\rangle\langle   \psi_{x, y}| \nonumber \\
                        &\;\;\;+     |3 \rangle\langle 3|\otimes |
                          \psi_{x+1, y+1}\rangle\langle   \psi_{x, y}| \Big],
\end{align}
where $|\psi_{x,y} \rangle =|\psi_x \rangle\otimes |\psi_y \rangle$.
It is well known that the operation $[\hat{S}_{(x,y)} [\hat{G} \otimes
\hat{{\mathbbm 1}}]]^t$ results in maximal spread of the probability
distribution only for the very specific initial state 
\begin{equation}
  |\Psi_{\rm in}^4\rangle = \frac{1}{2}\left(|0\rangle -|1\rangle -|2\rangle +
                                    |3\rangle\right)\otimes |\psi_{0, 0}\rangle,
  \label{eq:psi4}
\end{equation}
whereas the walk is localized at the origin for any other
case\,\cite{IKN04, SKJ08}. Choosing $|\Psi_{\rm in}^4\rangle$ and evolving it
for $t$ steps one finds 
\begin{eqnarray}
  |\Psi^4(t)\rangle&=&\sum_{x=-t}^t \sum_{y=-t}^t
    [ \mathcal{A}_{(x, y),t}|0 \rangle +\mathcal{B}_{(x, y),t}|1 \rangle \nonumber\\
    &&\;\;\;+ \mathcal{C}_{(x, y),t}|2 \rangle +\mathcal{D}_{(x, y),t}|3 \rangle]
    \otimes|\psi_{(x, y)}\rangle
\end{eqnarray}
where $\mathcal{A}_{(x, y),t}$, $\mathcal{B}_{(x, y),t}$,
$\mathcal{C}_{(x, y),t}$ and $\mathcal{D}_{(x, y),t}$ are given by the
iterative relations
\begin{subequations}
  \label{eq:4s_iter}
  \begin{eqnarray}
    \mathcal{A}_{(x, y),t} &=& \frac{1}{2} [-\mathcal{A}_{(x+1, y+1),t-1} +  \mathcal{B}_{(x+1, y+1),t-1} \nonumber \\
    && + \mathcal{C}_{(x+1, y+1),t-1} +  \mathcal{D}_{(x+1, y+1),t-1} ] \\
    \mathcal{B}_{(x, y),t} &=& \frac{1}{2} [\mathcal{A}_{(x+1, y-1),t-1} -  \mathcal{B}_{(x+1, y-1),t-1}  \nonumber \\
    &&+ \mathcal{C}_{(x-1, y-1),t-1} +  \mathcal{D}_{(x+1, y-1),t-1} ] \\
    \mathcal{C}_{(x, y),t} &=& \frac{1}{2} [\mathcal{A}_{(x-1, y+1),t-1} +  \mathcal{B}_{(x-1, y+1),t-1} \nonumber \\
    && - \mathcal{C}_{(x-1, y+1),t-1}  + \mathcal{D}_{(x-1, y+1),t-1} ]\\
    \mathcal{D}_{(x, y),t} &=& \frac{1}{2} [\mathcal{A}_{(x-1, y-1),t-1} +  \mathcal{B}_{(x-1, y-1),t-1} \nonumber \\
    && + \mathcal{C}_{(x-1, y-1),t-1} -  \mathcal{D}_{(x-1, y-1),t-1} ].
  \end{eqnarray}
\end{subequations}
This results in the probability distribution
\begin{align}
  \label{eq:4sp}
  P_{4s} = \sum_{x =-t}^t \sum_{y =-t}^t  
            \Big[|&\mathcal{A}_{(x, y),t}|^2 +|\mathcal{B}_{(x, y),t}|^2 \nonumber \\
         &+ |\mathcal{C}_{(x, y),t}|^2+|\mathcal{D}_{(x, y),t}|^2\Big]
\end{align}
which is shown in Fig.\,\ref{fig:1} for $t=25$.

\subsection{Alternate walk}
\label{2sqw1}

Very recently a 2D quantum walk was suggested which used only a
two-state particle which walks first only along the $x$-axis followed by
a step along the $y$-axis \cite{FGB11}. This walk can results in the same probability distribution as the Grover walk and its evolution is given by
\begin{equation}
  |\Psi_1\rangle = (\hat{S}_{(0,y)}
  [\hat{B}(\theta) \otimes \hat{{\mathbbm 1}}])(\hat{S}_{(x, 0)}
  [\hat{B}(\theta) \otimes \hat{{\mathbbm 1}}])|\Psi_{\rm in} \rangle,
\end{equation}
where
\begin{subequations}
\begin{eqnarray}
  \label{sx}
  \hat{S}_{(x, 0)} &\equiv&   \sum_{x, y}  \big[  |0 \rangle\langle
  0|\otimes|\psi_{x-1, y}\rangle\langle   \psi_{x, y}|  \nonumber \\
  &&\;\;\;\; +  |1 \rangle\langle
  1|\otimes |\psi_{x+1, y}\rangle\langle \psi_{x, y}| \big ]  \\
  \label{sy}
  \hat{S}_{(0, y)} &\equiv&   \sum_{x, y}  \big[  |0 \rangle\langle
  0|\otimes|\psi_{x, y-1}\rangle\langle   \psi_{x, y}| \nonumber \\
  &&\;\;\;\;  +  |1 \rangle\langle
  1|\otimes |\psi_{x, y+1}\rangle\langle \psi_{x, y}| \big].
\end{eqnarray}
\end{subequations}
 Using a coin operation with $\theta = \pi/4$, the state of the walk after $t$ steps can then be calculated as 
\begin{eqnarray}
  |\Psi(t)\rangle &=& \hat{W}(\pi/4)^t |\Psi_{\rm in}\rangle \nonumber \\
  &=& \sum_{x=-t}^t \sum_{y=-t}^t \left[ \mathcal{A}_{(x, y),t}|0 \rangle
    +\mathcal{B}_{(x, y),t}|1 \rangle \right ]\otimes |\psi_{(x,
    y)}\rangle\nonumber,\\
\label{eq:lr}
\end{eqnarray}
where $\hat{W}(\pi/4) = (\hat{S}_{(0,y)} [\hat{B}(\pi/4) \otimes
\hat{{\mathbbm 1}}])(\hat{S}_{(x, 0)} [\hat{B}(\pi/4) \otimes
\hat{{\mathbbm 1}}])$, and $\mathcal{A}_{(x, y),t}$ and
$\mathcal{B}_{(x, y),t}$ are given by the coupled iterative relations
\begin{subequations}
\label{eq:iter1}
\begin{align}
\mathcal{A}_{(x, y),t} = \frac{1}{2} \Big[&  \mathcal{A}_{(x+1, y+1),t-1} + \mathcal{A}_{(x-1, y+1),t-1} \nonumber \\
& + \mathcal{B}_{(x+1, y+1),t-1} -  \mathcal{B}_{(x-1, y+1),t-1} \Big]\\
\mathcal{B}_{(x, y),t} = \frac{1}{2} \Big[& \mathcal{A}_{(x+1, y-1),t-1} - \mathcal{A}_{(x-1, y-1),t-1} \nonumber \\ 
& + \mathcal{B}_{(x+1, y-1),t-1} + \mathcal{B}_{(x-1, y-1),t-1} \Big].
\end{align}
\end{subequations}
The resulting probability distribution is then
\begin{equation}
\label{eq:2sp}
P_{2s} = \sum_{x =-t}^t \sum_{y =-t}^t \left[|\mathcal{A}_{(x,
    y),t}|^2 + |\mathcal{B}_{(x, y),t}|^2 \right],
\end{equation} 
which for the initial state $|\Psi_{\rm
  in}\rangle = \frac{1}{\sqrt 2}(|0\rangle + i |1\rangle)\otimes
|\psi_{0,0}\rangle$ gives the same probability distribution as the four-state 
Grover walk (see Fig.\,\ref{fig:1}). 

\subsection{Pauli walk}
\label{2sqw2}

A further scheme to implement a 2D quantum walk using only a
two-state particle can be constructed using different Pauli basis
states as translational states for the two axis. For convenience we can
choose the eigenstates of the Pauli operator
$\hat{\sigma}_3= \begin{bmatrix} 1 & ~~0\\ 0 & -1 \end{bmatrix}$,
$|0\rangle$ and $|1\rangle$ as basis states for $x-$axis and
eigenstates of $\hat{\sigma}_1 = \begin{bmatrix} 0 &
  1\\ 1 & 0 \end{bmatrix}$, $|+\rangle = \frac{1}{\sqrt 2}(|0\rangle +
|1\rangle)$ and $|-\rangle = \frac{1}{\sqrt 2}(|0\rangle - |1\rangle)$
as basis states for $y-$axis\,\cite{CBS10}, which also implies that
$|0\rangle = \frac{1}{\sqrt 2}(|+\rangle + |-\rangle)$ and $|1\rangle
= \frac{1}{\sqrt 2}(|+\rangle - |-\rangle)$. In this scheme a coin
operation is not necessary and each step of the walk can be implemented by
the operation
\begin{equation}
   \hat{S}_{\sigma_3} \equiv \hat{S}_{(x, 0)}
\end{equation}
followed by the operation
\begin{align}
\hat{S}_{\sigma_1} \equiv   \sum_{x, y}  \Big [& |+ \rangle\langle
+|\otimes|\psi_{x, y-1}\rangle\langle   \psi_{x, y}| \nonumber \\
&+ |- \rangle\langle-|\otimes |\psi_{x, y+1}\rangle\langle \psi_{x, y}|\Big].
\label{ssigma}
\end{align}
The state after $t$ steps of quantum walk is then given by
\begin{eqnarray}
|\Psi_t\rangle &=&[\hat{S}_{\sigma_1}\hat{S}_{\sigma_3}]^t |\Psi_{\rm in}\rangle
                \nonumber \\
&=& \sum_{x =-t}^t \sum_{y =-t}^t \left [\mathcal{A}_{(x, y),t}|0
  \rangle +\mathcal{B}_{(x, y),t}|1\rangle \right ]\otimes|\psi_{x,
  y}\rangle,\nonumber\\
\label{eq:lr2}
\end{eqnarray}
where $\mathcal{A}_{(x, y),t}$ and $\mathcal{B}_{(x,y),t}$ are given by the
coupled iterative relations
\begin{subequations}
\label{eq:iter}
\begin{align}
\mathcal{A}_{(x, y),t} = \frac{1}{2} \Big[&\mathcal{A}_{(x+1, y+1),t-1} + \mathcal{B}_{(x-1, y+1),t-1}  \nonumber \\
& + \mathcal{A}_{(x+1, y-1),t-1} -  \mathcal{B}_{(x-1, y-1),t-1} \Big] \\
\mathcal{B}_{(x, y),t} =\frac{1}{2} \Big[&\mathcal{B}_{(x-1, y+1),t-1} +  \mathcal{A}_{(x+1, y+1),t-1} \nonumber \\
& + \mathcal{B}_{(x-1, y-1),t-1} -  \mathcal{A}_{(x+1, y-1),t-1} \Big].
\end{align}
\end{subequations}
The probability distribution 
\begin{equation}
\label{eq:2spsigma}
 P_{2s\sigma} = \sum_{x =-t}^t \sum_{y =-t}^t  
             \left[|\mathcal{A}_{(x, y),t}|^2 + |\mathcal{B}_{(x, y),t}|^2 \right]
\end{equation}
is again equivalent to the distribution obtained using the Grover walk and therefore also to the alternative
walk for the initial state $|\Psi_{\rm in}\rangle =
\frac{1}{\sqrt 2}(|0\rangle + i |1\rangle)\otimes |\psi_{0,0}\rangle$ (see Fig.\,\ref{fig:1}). 

While the shift operator for the Grover walk is defined by a single
operation, experimentally it has to be implemented as a two shift
operations. For example, to shift the state $|0\rangle$ from $(x, y)$
to $(x-1, y-1)$, it has first to be shifted along one axis followed by
the other, very similarly to the way it is done in the two-state
quantum walk schemes.  Therefore, a two-state quantum walk in 2D has many 
advantages over a four-state quantum walk. The two-state walk using different 
Pauli basis states for the different axes has the further advantage of not requiring a coin 
operation at all, making the experimental task even simpler in physical systems where
access to different Pauli basis states as translational states is available.  One can, of course, also consider including a coin operation in the Pauli walk, which would result in
a different probability distributions \cite{Cha12}.

For general initial states, quantum walks in 2D
with a coin operation $\in U(2)$ can result in a many non-localized
probability distribution in position space. This is a further difference to
the Grover walk which is very specific with respect to the initial
state of the four-state particle and the coin operation.

\section{Measurement Induced Disturbance}
\label{qmid}
Quantifying non-classical correlations inherent in a certain state is
currently one of the most actively studied topics in physics (see for
example \cite{MBC11}). While many of the suggested methods involve
optimization, making them computationally hard, Luo\,\cite{L08}
recently proposed a computable measure that avoids this complication:
if one considers a bipartite state $\rho$ living in the Hilbert space
${\cal H}_A \otimes {\cal H}_B$, one can define a reasonable measure of
the total correlations between the systems $A$ and $B$ using the mutual
information
\begin{equation}
I(\rho) = S(\rho_A) + S(\rho_B) - S(\rho),
\label{eq:I}
\end{equation}
where $S(\cdot)$ denotes von Neumann entropy and $\rho_A$ and $\rho_B$
are the respective reduced density matrices. If $\rho_A = \sum_j p_A^j\Pi_A^j$
and $\rho_B = \sum_j p_B^j\Pi_B^j$, then the measurement induced by
the spectral components of the reduced states is
\begin{equation}
\label{eq:Pi}
\Pi(\rho) \equiv \sum_{j,k} \Pi_A^j \otimes \Pi_B^k \rho
\Pi_A^j \otimes \Pi_B^k.
\end{equation}
Given that $I[\Pi(\rho)]$ is a good measure of {\it classical}
correlations in $\rho$, one may consider a measure for quantum
correlations defined by the so-called Measurement Induced Disturbance
(MID)  \cite{L08}
\begin{equation}
\label{eq:Q}
Q(\rho) = I(\rho) - I[\Pi(\rho)].
\end{equation}
MID does not involve any optimization over local measurements
and can be seen as a loose upper bound on quantum discord \cite{OZ01}. At the same time it is known to capture most of the detailed trends in the
behaviour of quantum correlations during quantum walks\,\cite{RSC11}. Therefore, we will use MID ($Q(\rho)$) in
the following to quantify quantum correlations for the different 2D
quantum walk evolutions.

\begin{figure}[tb]
	\includegraphics[width=8.6cm]{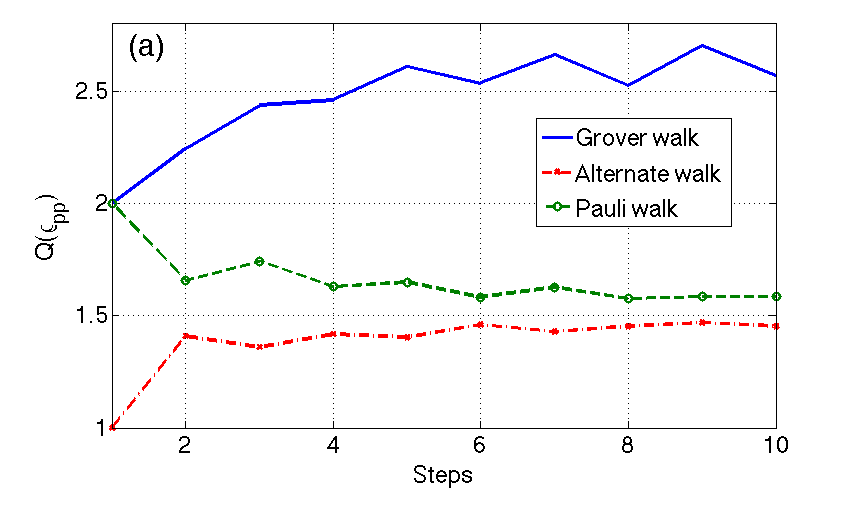}
	\label{fig:2a} \\
	\includegraphics[width=8.6cm]{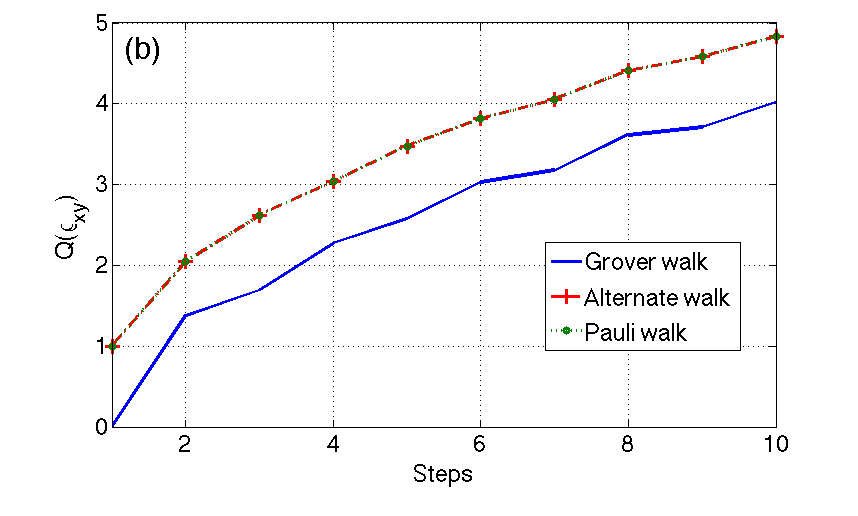}
	\label{fig:2b} 
	\caption{(Color online) Quantum correlations $Q(\rho_{pp})$ and $Q(\rho_{xy})$ for the different
	schemes in the absence of noise.\label{fig:2}}
\end{figure}
Despite having the same probability distributions in the absence of
noise, the MIDs for the four-state walk and the two-state walks differ. In
Fig.\,\ref{fig:2}(a) we show the MID between the particle and the position
degree of freedom, $Q(\rho_{pp})$ for all three walks and find that it is 
significantly higher for the Grover walk. Where $\rho_{pp}$ is $\tilde{\rho}_{4s}(t)$
for Grover walk, $\tilde{\rho}_{2s}(t)$ for alternate walk and $\tilde{\rho}_{2s\sigma}(t)$ for the Pauli walk. However, due to the difference in the size of the particles degree of freedom for the Grover and the two-state walks, a direct comparison of the quantum correlations $Q(\rho_{pp})$ does 
not make sense. Among the two-state schemes on the other hand, we see 
that the Pauli walk has a larger $Q(\rho_{pp})$ in comparison to  the alternate 
walk. 

A fair comparison between all systems can be made by looking at the quantum correlations 
generated between the two spatial dimensions $x$ and $y$, $Q(\rho_{xy})$ (see Fig.\,\ref{fig:2}(b)). 
$\rho_{xy}$ is obtained by tracing out the particle degree of freedom from from complete density matrix [$\tilde{\rho}_{4s}(t)$, $\tilde{\rho}_{2s}(t)$ and $\tilde{\rho}_{2s\sigma}(t)$] comprising of the particle and the position space. We find that  $Q(\rho_{xy})$ is identical for both two-state schemes and exceeding 
the Grover walk result.  This behaviour is similar to the one described in 
Refs. \cite{FGB11, FGM11}, where the entanglement created during the Grover 
walk was compared with the alternate walk using the negativity of the partial transpose, 
in its generalization for higher-dimensional systems  \cite{LCO03, LKS11}.

\section{Decoherence}
\label{decoqw}

The effects of noise on 1D quantum walks has been widely
studied \cite{Ken07, CSB07, BSC08, SBC10, RSC11} but the implications
in 2D settings are less well known \cite{OPD06, GAM09, Amp12}. In
particular no study has been done on either of the two-state schemes
presented in the previous section and we therefore now compare their
decoherence properties to the Grover walk, using a state-flip and a
depolarizing channel as noise models. We show that this leads to
differing probability distributions and has an effect on the amounts of
quantum correlations as well.

\subsection{State flip noise}
\label{2sf}
\subsubsection{Grover walk}
\label{4sf}
For a two-state particle, state-flip noise simply induces a bit flip
[$\sigma_1 = \begin{bmatrix} 0 & 1\\ 1 & 0 \end{bmatrix}$] but for the
Grover walk, the state-flip noise on the four basis states can change one state to
23 other possible permutations. Therefore, the density
matrix after $t$ steps in the presence of a state-flip noise channel
can be written as
\begin{eqnarray}
\label{4s23}
   \hat{\rho}_{4s}(t) &=& \frac{p}{k} 
                         \left[\sum_{i =1}^k\hat{f}_i\hat{S}_4\hat{\rho}_{4s}(t-1)
                               \hat{S}_4{^\dagger} \hat{f}_i{^\dagger} \right] 
                         \nonumber \\
                     & &+(1-p)[\hat{S}_4\hat{\rho}_{4s}(t-1)\hat{S}_4{^\dagger}],
\end{eqnarray}
where $p$ is the noise level, $S_4= \hat{S}_{(x,y)} [\hat{G} \otimes
\hat{{\mathbbm 1}}]$, and the $\hat{f}_{i}$ are
the state-flip operations.  
\begin{figure}[tb]
\bc 
\subfigure[]{\includegraphics[width=4.2cm]{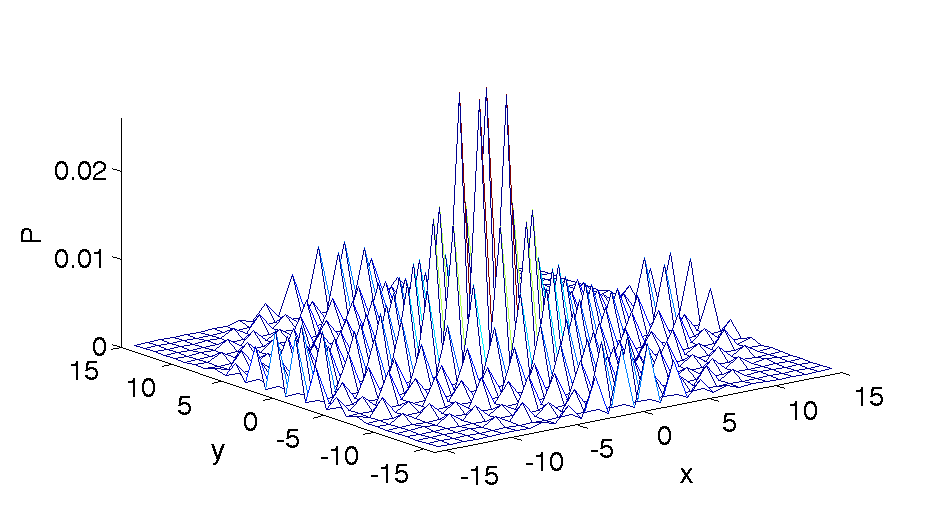}
\label{fig:3a}}
\hskip -0.1in
\subfigure[]{\includegraphics[width=4.2cm]{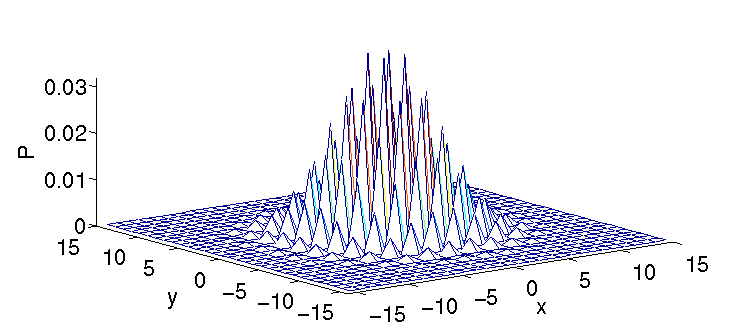}
\label{fig:3b}}
\ec
\caption{(Color online) Probability distribution of
  the Grover walk when subjected to a different state-flip noise level with
  $k=23$ after 15 steps. (a) and (b) are for noise levels, $p=0.1$ and $p=0.9$, respectively.}
\end{figure}
\begin{figure}[ht]
\includegraphics[width=8.6cm]{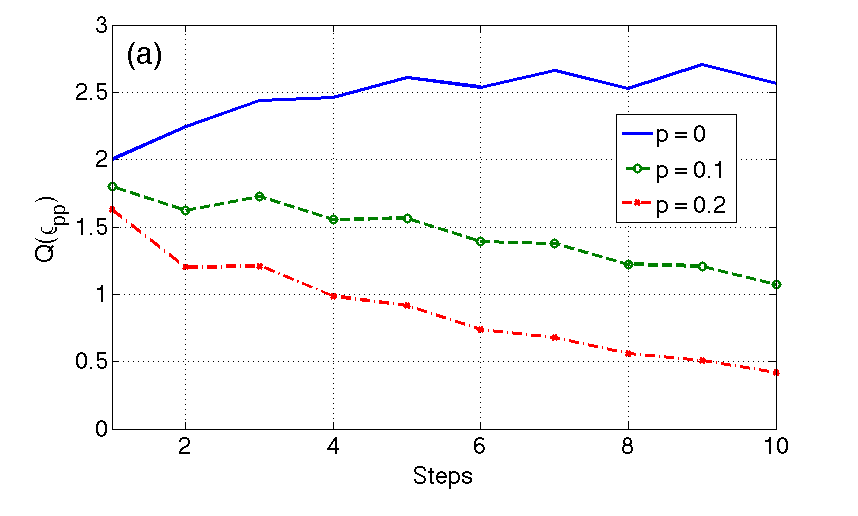}\\
\includegraphics[width=8.6cm]{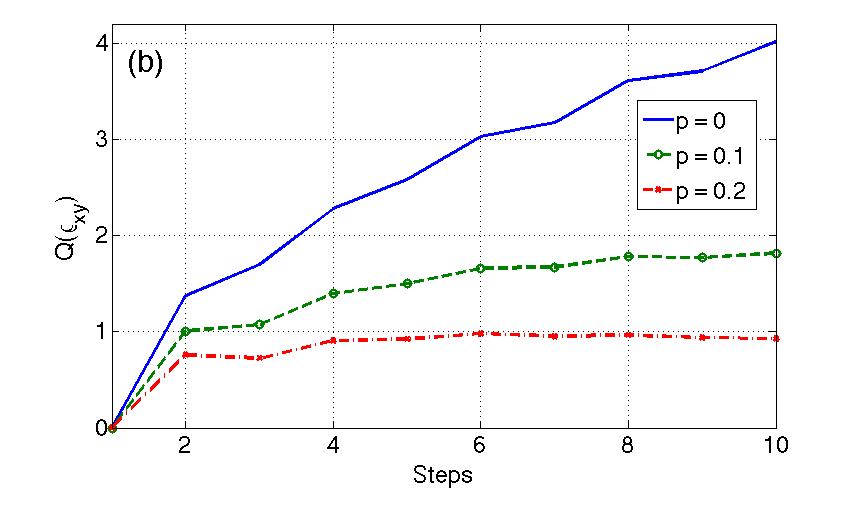}
\caption{(Color online) Quantum correlations created by the Grover walk in the presence of a noise channel including all possible state flips ($k=23$).\label{fig:4}}
\end{figure}
For a noisy channel with all the 23
possible flips one has $k=23$ and in Figs.\,\ref{fig:3a} and
\ref{fig:3b} we show the probability distribution of the Grover
for weak ($p=0.1$) and strong ($p=0.9$) noise levels after 15 step. Compared to the distribution in the absence of noise (see Fig.~\ref{fig:1}) a progressive reduction in the quantum spread is clearly visible. Note that for
$p=1$ the walk corresponds to a fully classical evolution. 

In Fig.\,\ref{fig:4}(a) we show the quantum correlations 
between the particle state with the position space, $Q(\rho_{pp})$,  
as a function of number of steps $t$. With increasing noise level, a decrease in  $Q(\rho_{pp})$ is seen, whereas for the
 quantum correlations between the $x$ and $y$ spatial 
dimensions, $Q(\rho_{xy})$, the same amount of noise mainly leads to a decrease in the positive slope (see Fig.\,\ref{fig:4}(b)).

\subsubsection{Two-state walks}
\label{ssn}
The evolution of each step of the two-state quantum walk comprises of
a move along one axis followed a move along the other. Therefore, the
walk can be subjected to a noise channel after evolution along each
axis or after each full step of the walk. In the first case, the noise
level $p^{\prime}=\frac{p}{2}$ is applied two times during each step
in oder to be equivalent to the application of a noise of strength $p$
in the second case. For the alternate walk the evolution of the
density matrix with a bit-flip noise channel applied after evolution
along each axis is then given by
\begin{subequations}
\begin{align}
\label{eq:14a}
\hat{\rho}_{2s}^{\prime}(t) = &\frac{p}{2} \left [ \hat{\sigma}_1 \hat{S}_x \hat{\rho}_{2s}(t-1) \hat{S}_x^{\dagger} \hat{\sigma}_1^{\dagger} \right ] \nonumber \\
&+ \left (1-\frac{p}{2}\right )\left[\hat{S}_x \hat{\rho}_{2s}(t-1) \hat{S}_x^{\dagger} \right ] \\
\label{eq:14b}
\hat{\rho}_{2s}(t) =& \frac{p}{2} \left [ \hat{\sigma}_1\hat{S}_y \hat{\rho}_{2s}^{\prime}(t) \hat{S}_y^{\dagger}\hat{\sigma}_1^{\dagger} \right ] \nonumber \\
&+ \left (1-\frac{p}{2} \right )\left[\hat{S}_y \hat{\rho}_{2s}^{\prime}(t) \hat{S}_y^{\dagger} \right ],
\end{align}
\end{subequations}
where  $\hat{\sigma}_1 = \begin{bmatrix}0   &  1  \\
  1 & 0
\end{bmatrix}\otimes  \hat{{\mathbbm 1}}$, $\hat{S}_y = \hat{S}_{(0,y)} [\hat{B}(\theta) \otimes  \hat{{\mathbbm 1}}]$, and $\hat{S}_x=\hat{S}_{(x, 0)} [\hat{B}(\theta) \otimes  \hat{{\mathbbm 1}}]$.
\par
Similarly, the density matrix with a bit-flip noise
applied after the evolution along each axis for the Pauli walk is
given by
\begin{subequations}
\begin{align}
\label{eq:15a}
\hat{\rho}_{2s\sigma}^{\prime}(t) = &\frac{p}{2} \left [\hat{\sigma}_1 \hat{S}_{\sigma_3} \hat{\rho}_{2s\sigma}(t-1) \hat{S}_{\sigma_3}^{\dagger} \hat{\sigma}_1^{\dagger} \right ] \nonumber \\
&+ \left (1-\frac{p}{2}\right )\left[\hat{S}_{\sigma_3} \hat{\rho}_{2s\sigma}(t-1) \hat{S}_{\sigma_3}^{\dagger} \right ] \\
\label{eq:15b}
\hat{\rho}_{2s\sigma}(t) =& \frac{p}{2} \left [ \hat{\sigma}_1 \hat{S}_{\sigma_1} \hat{\rho}_{2s\sigma}^{\prime}(t) \hat{S}_{\sigma_1}^{\dagger}\hat{\sigma}_1^{\dagger} \right ] \nonumber \\
&+ \left (1-\frac{p}{2} \right )\left[\hat{S}_{\sigma_1} \hat{\rho}_{2s\sigma}^{\prime}(t) \hat{S}_{\sigma_1}^{\dagger} \right ].
\end{align}
\end{subequations}
\begin{figure}[tb]
\bc 
\subfigure[Alternate walk]{\includegraphics[width=4.25cm]{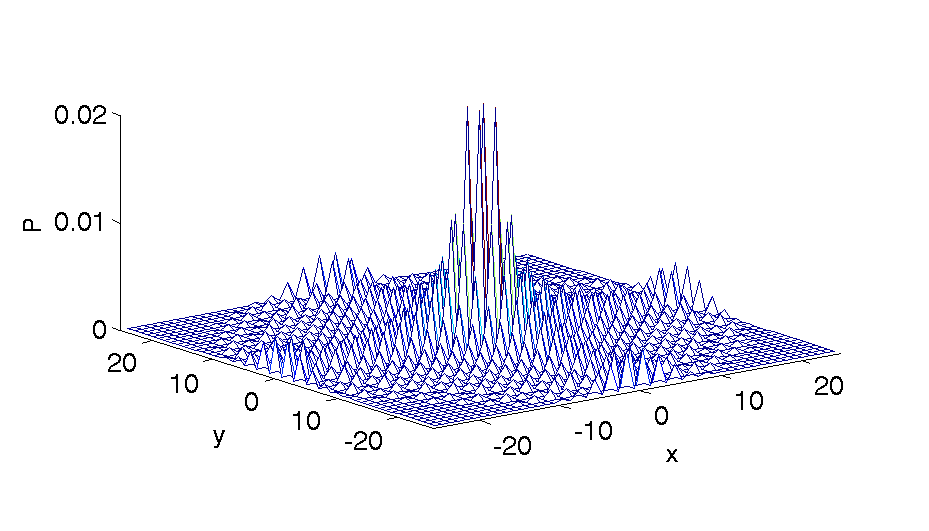} 
\label{fig:5a}}
\hskip -0.1in
\subfigure[Alternate walk]{\includegraphics[width=4.25cm]{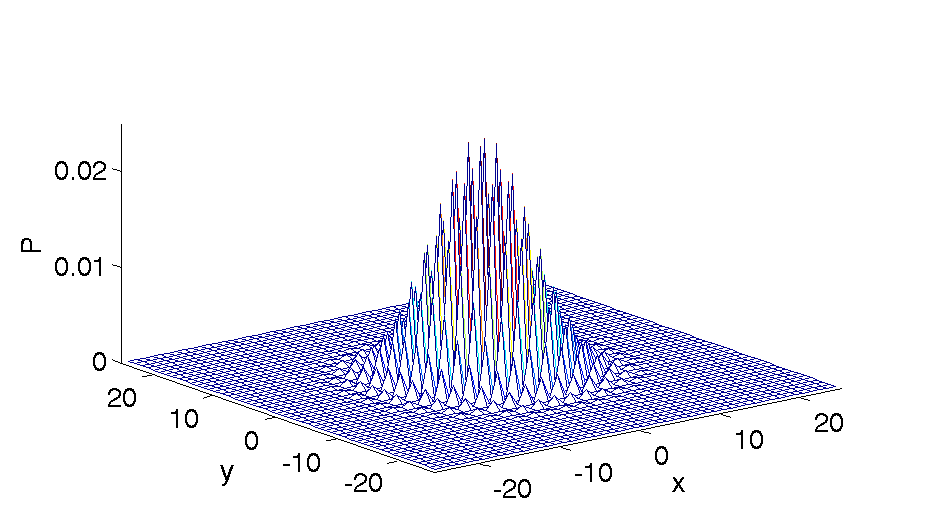}
\label{fig:5b}}
\subfigure[Pauli walk]{\includegraphics[width=4.25cm]{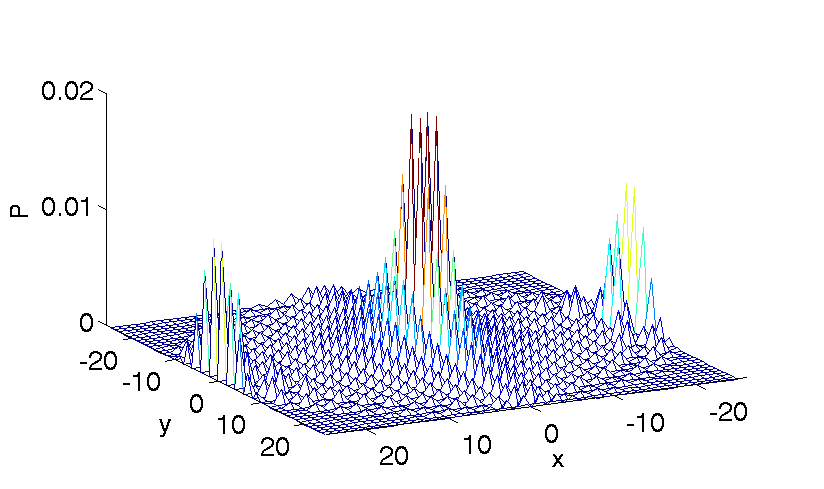}
\label{fig:5c}}
\hskip -0.1in
\subfigure[Pauli walk]{\includegraphics[width=4.25cm]{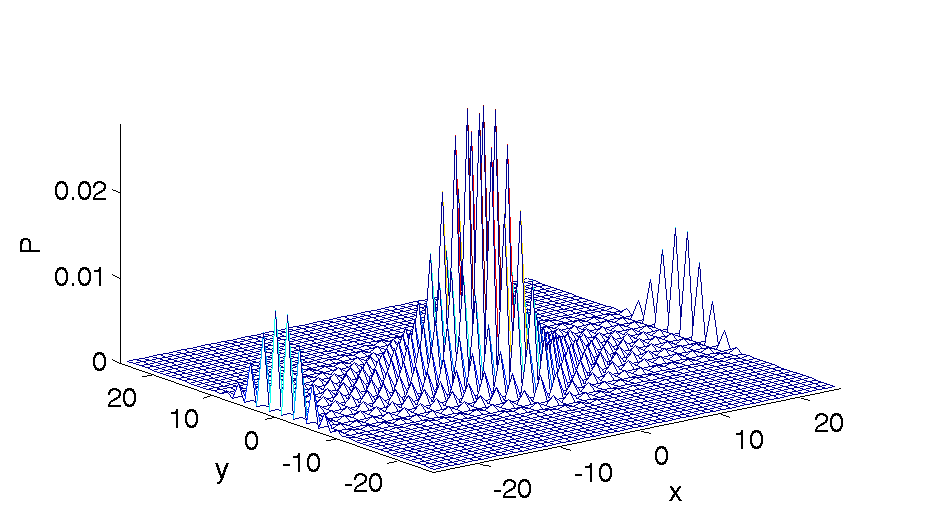}
\label{fig:5d}}
\ec
\caption{(Color online) Probability distributions of the two-state walk with the bit-flip noise applied after evolution in each direction.  The strength of the noise is $p/2=0.05$ in (a) and (c) and $p/2=0.45$ in (c) and (d) and the evolution was carried out for 25 step each time.}
  \label{fig:5}
\end{figure}
\begin{figure}[ht]
\subfigure[Alternate walk]{\includegraphics[width=8.6cm]{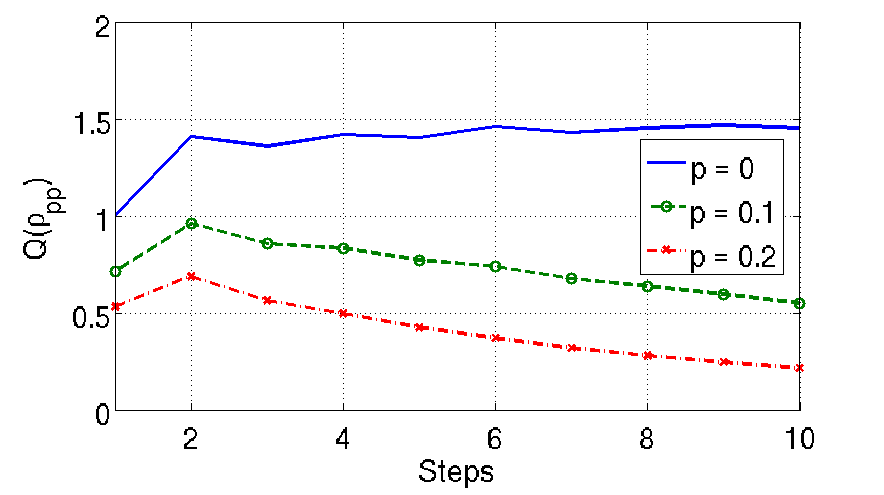}
\label{fig:6a}}
\subfigure[Pauli Walk]{\includegraphics[width=8.6cm]{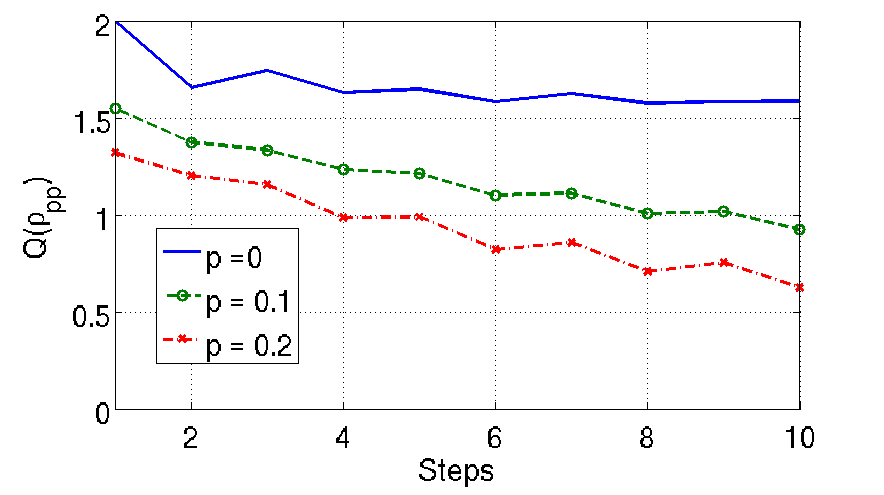}
\label{fig:6b}} 
\caption{(Color online) Particle-position quantum correlations created by the two-state walks for different bit-flip noise levels applied after evolution along each axis.}
\label{fig:6}
\end{figure}
\begin{figure}[ht]
\subfigure[Alternate walk]{\includegraphics[width=8.6cm]{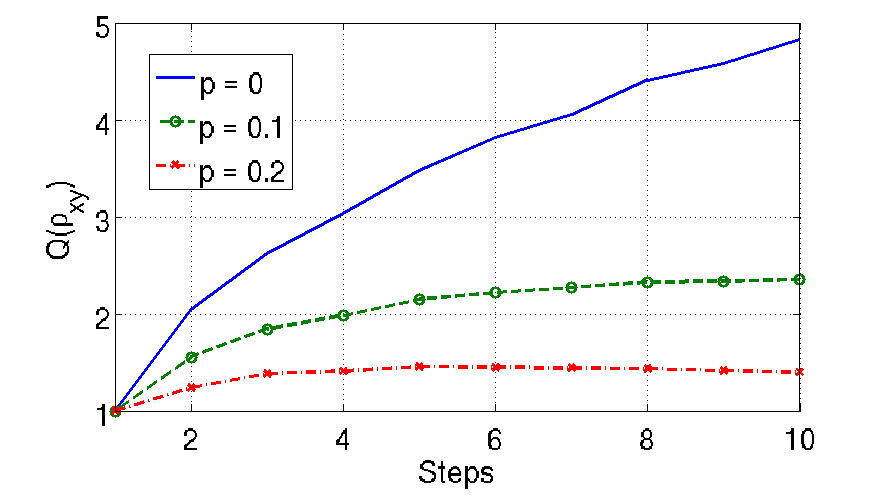}
\label{fig:7a}}
\subfigure[Pauli walk]{\includegraphics[width=8.6cm]{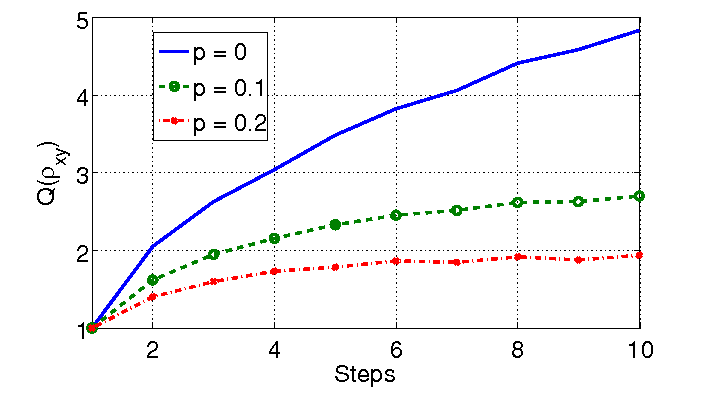}
\label{fig:7b}}
\caption{(Color online) Quantum correlations between the $x$ and $y$ spatial dimensions created by the two-state walks for different bit-flip noise levels applied after evolution along each axis}
\label{fig:7}
\end{figure}

In Figs.\,\ref{fig:5a} and \ref{fig:5b} the probability distributions
for the alternate walk after 25 steps of noisy evolution with $p=0.1$
and $p=0.9$ are shown and Figs.\,\ref{fig:5c} and \ref{fig:5d} show
the same for the Pauli walk.  It can be seen that the bit-flip
noise channel acts symmetrically on both axes for the alternate walk,
but asymmetrically on the Pauli walk. This is due to the fact that the
bit-flip noise applied along the axis in which the $\sigma_1$ Pauli basis is
used leaves the state unchanged. A completely classical evolution is recovered for $p=1$ ($p^{\prime}=0.5$) and an evolution with $p=2$ is equivalent to one with $p=0$. 

Evolving the density matrix and calculating the MID for a noiseless evolution ($p=0$), one can see from Fig.~\ref{fig:6} that the initial difference in  $Q(\rho_{pp})$ between the Pauli walk and the alternate walk decreases during the evolution and eventually both values settle around 1.5. For a noisy evolution, however, the initial difference in $Q(\rho_{pp})$ 
does not decrease over time and we find a higher value for the Pauli walk compared to the alternate walk. Similarly, careful examination of  Fig.\,\ref{fig:7} shows that the $Q(\rho_{xy})$ for the alternate walk and the Pauli walk are identical in the absence of noise, but differ for noisy evolution, with the alternate walk being affected stronger than the Pauli walk.

\begin{figure}[ht]
\bc 
\subfigure[Alternate walk]{\includegraphics[width=8.6cm]{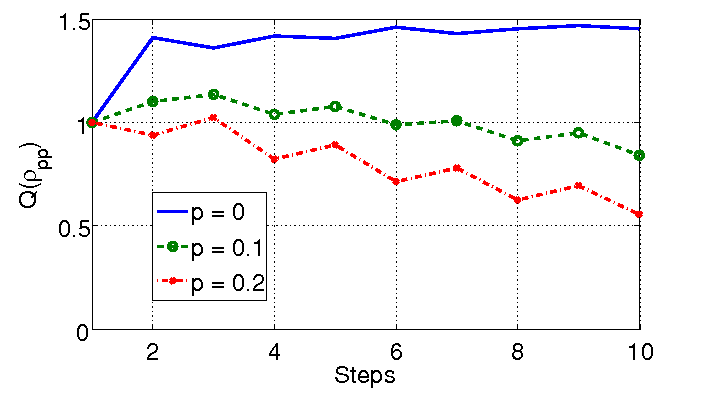}
\label{fig:8b}}
\subfigure[Pauli walk]{\includegraphics[width=8.6cm]{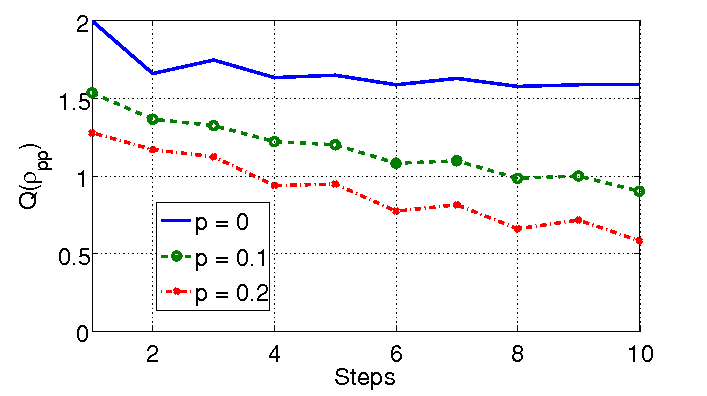}
\label{fig:8c}}
\ec
\caption{ (Color online) Particle-position quantum correlations created by the two-state walks for different bit-flip noise levels applied after evolution of one complete step.\label{fig:8}}
\end{figure}

\begin{figure}[ht]
\bc 
\subfigure[Alternate walk]{\includegraphics[width=8.6cm]{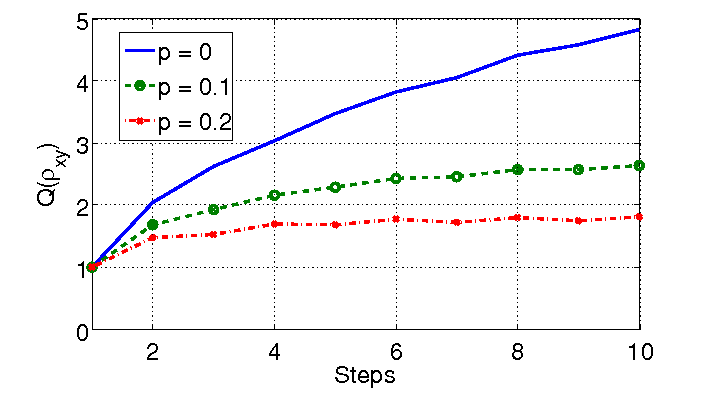}
\label{fig:9b}}
\subfigure[Pauli walk]{\includegraphics[width=8.6cm]{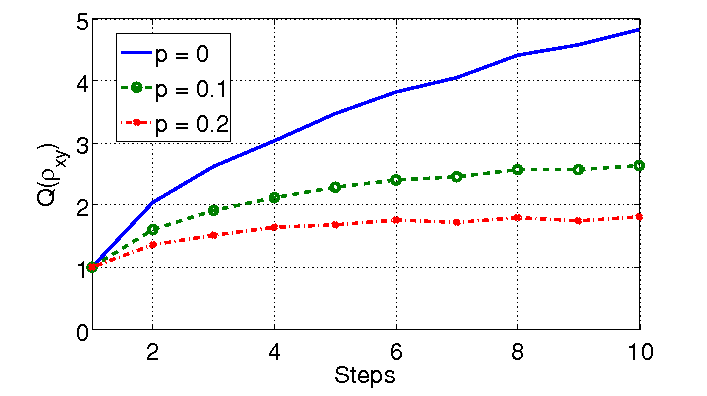}
\label{fig:9c}}
\ec
\caption{ (Color online) Quantum correlations between the $x$ and $y$ spatial dimensions created by the two-state walks for different bit-flip noise levels applied after evolution of one complete step.\label{fig:9}}
\end{figure}

The density matrix for the second case, that is, with a noisy channel applied only once
after one full step of walk evolution, for both two-state walks is given by
\begin{subequations}
\begin{align}
\label{eq:16a}
\hat{\rho}_{2s}(t) &= p \left [ \hat{\sigma}_1\hat{Q}(t-1)\hat{\sigma}_1^{\dagger} \right ]
+ (1-p) \hat{Q}(t-1)  \\
\label{eq:16b}
\hat{\rho}_{2s\sigma}(t) &= p \left [ \hat{\sigma}_1   \hat{Q}_{\sigma}(t-1)   \hat{\sigma}_1^{\dagger} \right ] 
+ (1-p) \hat{Q}_{\sigma}(t-1) 
\end{align}
\end{subequations}
where
\begin{subequations}
\begin{align}
\label{q1}
\hat{Q}(t-1) &= \hat{S}_y \hat{S}_x \hat{\rho}_{2s}(t-1) \hat{S}_x^{\dagger} \hat{S}_y^{\dagger} \\
\label{q2} 
\hat{Q}_{\sigma}(t-1)  &= \hat{S}_{\sigma_1} \hat{S}_{\sigma_3} \hat{\rho}_{2s\sigma}(t-1) \hat{S}_{\sigma_3}^{\dagger} \hat{S}_{\sigma_1}^{\dagger}.
\end{align}
\end{subequations}
In this case maximum decoherence and a completely classical evolution is obtained for $p=0.5$ and the evolution with $p=1$ is equivalent to the one with $p=0$. 
The obtained probability distributions are almost identical for both walks and differ only slightly from the ones obtained for the alternate walk with noise applied after evolution along each axis (see Fig.\,\ref{fig:5a} and \ref{fig:5b}). The correlation functions $Q(\rho_{pp})$ and $Q(\rho_{xy})$ behave very similar for both walks  (see Figs.~\ref{fig:8} and \ref{fig:9}) and we can conclude that the presence of bit-flip noise on both two-state walks when applied after a full step leads to equally strong decoherence.

\subsection{Depolarizing channel}
\label{sec:depol}

To describe depolarizing noise we use the standard model in which the density matrix of our two-state system
is replaced by a linear combination of a completely mixed and an unchanged state, 
\begin{equation}
\hat{\rho} = \frac{p}{3}
\left(\hat{\sigma}_{1}\hat{\rho} \hat{\sigma}_{1} +
  \hat{\sigma}_{2}\hat{\rho} \hat{\sigma}_{2}+
  \hat{\sigma}_{3}\hat{\rho} \hat{\sigma}_{3} \right ) +
(1-p)\hat{\rho}, 
\end{equation} 
where $\hat{\sigma}_1$, $\hat{\sigma}_2$ and $\hat{\sigma}_3$ are the
standard Pauli operators.  To be able to compare the effects of the
depolarizing channel on the Grover walk and the two-state
walks we will apply the noise only once after each
full step. 

\subsubsection{Grover walk}
\label{grodepol}

\begin{figure}[tb]
\includegraphics[width=8.4cm]{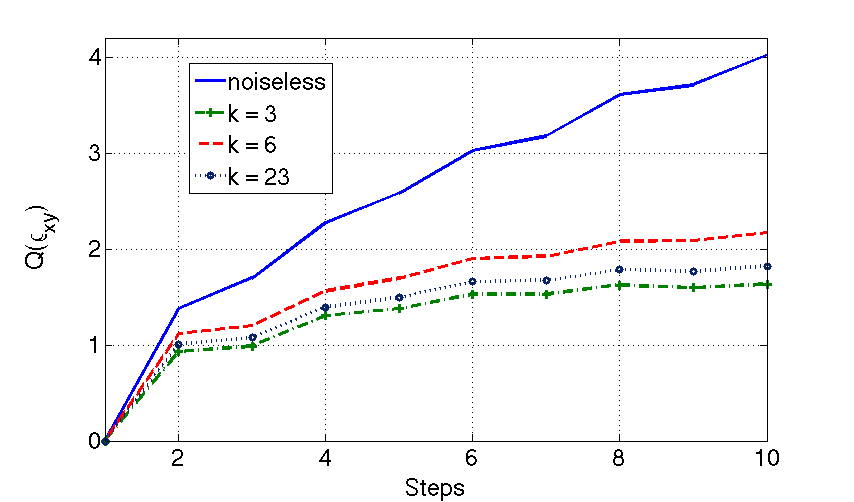}
\caption{(Color online) Quantum correlations between the $x$ and $y$ spatial dimensions for the Grover walk in the presence of a state-flip noise channel with $p=0.1$. The noise is modelled as state-flips including all possible flips ($k=23$), flips between only two of the basis states ($k=6$) and cyclic flips of all the four basis state ($k=3$)\label{fig:10}}
\end{figure}

For the four-state particle the depolarizing noise channel comprises of all possible state flips, phase flips and their combinations. State-flip noise alone leads to 23 possible changes in the four-state system and adding the phase-flip noise and all combinations of these two is unfortunately a task beyond current computational ability. Therefore let us first
briefly investigate the possibility of approximating the state flip
noise by restricting ourselves to only a subset of flips. One example would be a
noisy channel with only 6 possible flips ($k=6$) between two of the
four basis states
\bea \hat{f}_1 = \begin{bmatrix} 0 & 1 & 0
  & 0 \\ 1 & 0 & 0 & 0\\ 0 & 0 & 1 & 0\\ 0 & 0 & 0 & 1 \end{bmatrix}
\otimes \hat{{\mathbbm 1}} ~~;~~
\hat{f}_2 = \begin{bmatrix} 0 & 0 & 1 & 0 \\ 0 & 1 & 0 & 0\\ 1 & 0 & 0 & 0\\ 0 & 0 & 0 & 1 \end{bmatrix} \otimes \hat{\mathbbm 1}\nonumber \\
\hat{f}_3 = \begin{bmatrix} 0 & 0 & 0 & 1 \\ 0 & 1 & 0 & 0\\ 0 & 0 & 1
  & 0\\ 1 & 0 & 0 & 0 \end{bmatrix} \otimes \hat{\mathbbm 1} ~~;~~
\hat{f}_4 = \begin{bmatrix} 1 & 0 & 0 & 0 \\ 0 & 0 & 1 & 0\\ 0 & 1 & 0 & 0\\ 0 & 0 & 0 & 1 \end{bmatrix} \otimes \hat{\mathbbm 1} \nonumber \\
\hat{f}_5 = \begin{bmatrix} 1 & 0 & 0 & 0 \\ 0 & 0 & 0 & 1\\ 0 & 0 & 1
  & 0\\ 0 & 1 & 0 & 0 \end{bmatrix} \otimes \hat{\mathbbm 1} ~~;~~
\hat{f}_6 = \begin{bmatrix} 1 & 0 & 0 & 0 \\ 0 & 1 & 0 & 0\\ 0 & 0 & 0
  & 1\\ 0 & 0 & 1 & 0 \end{bmatrix} \otimes \hat{\mathbbm 1}
\label{6flips}
\eea 
and another a channel where only cyclic flips ($k=3$) of all the basis states can appear
\begin{equation}
  \hat{f}_1 = \hat{f} \otimes \hat{{\mathbbm
      1}} ~~;~~ \hat{f}_2 = \hat{f}^2\otimes \hat{{\mathbbm 1}} ~~;~~
  \hat{f}_3 = \hat{f}^3\otimes \hat{{\mathbbm 1}}
\label{3flips}
\end{equation}
with $\hat{f}= \begin{bmatrix} 0 & 1 & 0 & 0 \\ 0 & 0 & 1 & 0\\ 0 & 0
  & 0 & 1\\ 1 & 0 & 0 & 0 \end{bmatrix}$.  The probability
  distributions for these two approximations are visually
  very similar to the situation where all possible state-flips are taken into account ($k=23$) and in Fig.~\ref{fig:10}
we compare the results obtained for the  
  the $x-y$ spatial quantum correlations for a  noise level of $p=0.1$.

One can see that the spatial quantum correlations  are affected stronger by the $k=3$ than by the full $k=23$ flip noise. This implies that the two- and three state flips included in $k=23$ acts as reversals of cyclic flips, thereby reducing the effect of noise. Since the trends for the decrease of the quantum correlation are functionally similar for $k=3, 6$ and 23, we will use the model with $k=3$ cyclic flips as the state-flip noise channel for the Grover walk in this section. Similarly, taking into account the computational limitations and following the model adopted for the state-flip, we will use a cyclic phase-flips to model the phase-flip noise, 
\begin{equation}
\hat{r}_1 = \hat{r} \otimes \hat{\mathbbm 1}~~;~~ \hat{r}_2 =
\hat{r}^2\otimes \hat{\mathbbm 1} ~~;~~ \hat{r}_3 = \hat{r}^{3}\otimes
\hat{\mathbbm 1}, 
\end{equation} 
where $\hat{r} = \begin{bmatrix} 1 & 0 & 0 & 0 \\ 0 & \omega & 0 & 0
  \\ 0 & 0 & \omega^2 & 0 \\ 0 & 0 & 0 & \omega^3 \end{bmatrix}$ with
$\omega = e^{\frac{2\pi i}{4}}$.
\begin{figure}[tb]
\bc 
\subfigure[]{\includegraphics[width=4.2cm]{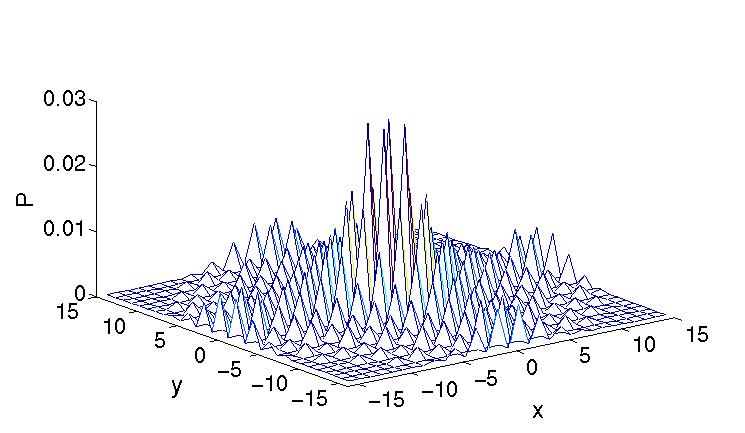}
\label{fig:11a}}
\hskip -0.1in
\subfigure[]{\includegraphics[width=4.2cm]{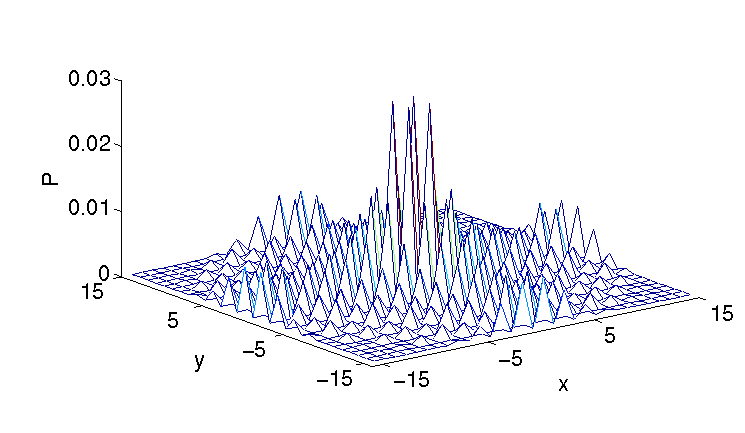}
\label{fig:11b}}
\ec
\caption{(Color online) Probability distribution of (a) the Grover
  walk and (b) the two-state walks when subjected to a depolarizing channel. For the Grover walk the channel is given by Eq.~\eqref{4sdepol}  and for all walks the noise level is $p=0.1$. The distribution is shown after 15 steps of evolution.}
\label{fig:11}
\end{figure}

\begin{figure}[tb]
\bc 
\subfigure[Grover walk]{\includegraphics[width=8.6cm]{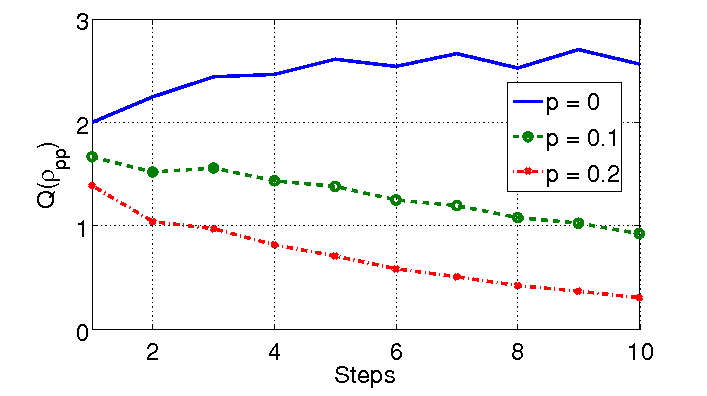}
\label{fig:12a}}
\subfigure[Grover walk]{\includegraphics[width=8.6cm]{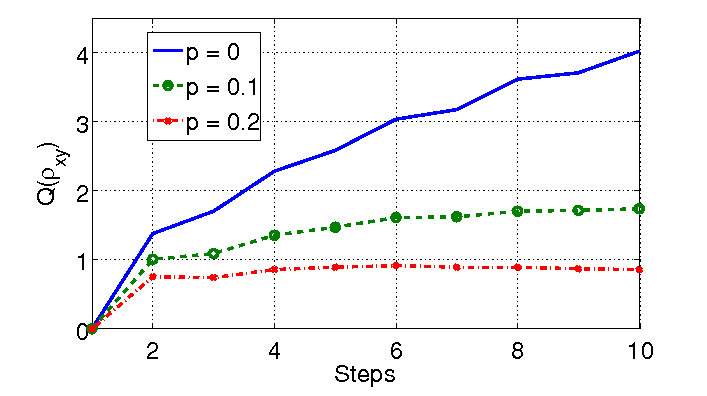}
\label{fig:12b}}
\ec
\caption{(Color online) Quantum correlations created by the Grover walk for different depolarizing noise levels.\label{fig:12}}
\end{figure}

Both these approximations for state-flip and phase-flip noise will makes the complex depolarization noise manageable for numerically treatment. The density matrix of the Grover walk can then be written as 
\begin{eqnarray}
\label{4sdepol}
\hat{\rho}_{4s}(t) &=& \frac{p}{15} \left [ \sum_{i =1}^{3} \hat{f}_i \hat{S}_4 \hat{\rho}_{4s}(t-1) \hat{S}_4{^\dagger} \hat{f}_i{^\dagger} \right ] \nonumber \\
&+& \frac{p}{15} \left [ \sum_{j=1}^{3} \hat{r}_j \hat{S}_4 \hat{\rho}_{4s}(t-1) \hat{S}_4{^\dagger} \hat{r}_j{^\dagger} \right ]\nonumber \\
&+& \frac{p}{15} \left [ \sum_{i=1}^{3} \sum_{j=1}^{3} \hat{r}_j \hat{f}_i \hat{S}_4 \hat{\rho}_{4s}(t-1) \hat{S}_4{^\dagger} \hat{f}_i^{\dagger} \hat{r}_j{^\dagger} \right ]\nonumber \\
& &+ (1-p)[\hat{S}_4 \hat{\rho}_{4s}(t-1) \hat{S}_4{^\dagger}]
\end{eqnarray}
and the probability distribution for this walk is shown in Fig.\,\ref{fig:11a} for $p=0.1$. The quantum correlations $Q(\rho_{pp})$ and $Q(\rho_{xy})$ are shown in Fig.\,\ref{fig:12}. With increasing noise level, a decrease in  $Q(\rho_{pp})$ is seen, whereas for the quantum correlations between the $x$ and $y$ spatial dimensions, $Q(\rho_{xy})$, the same amount of noise mainly leads to a decrease in the positive slope (see Fig.\,\ref{fig:12}).  From this we can conclude that the general trend in the quantum correlaltion due to state-flip noise (Fig.\,\ref{fig:4}) and depolarizing noise is the same but the effect is slightly stronger when including the depolarizing channel. 

\subsubsection{Two-state walks}
The depolarizing channels for the alternate walk and Pauli walk can be written
as
\begin{subequations}
  \begin{align}
    \label{2sdepol1}
    \hat{\rho}_{2s}(t) &= \frac{p}{3} \left [\sum_{i=1}^{3} \hat{\sigma}_i \hat{Q}(t-1) \hat{\sigma}_i^{\dagger} \right ] + (1-p) \hat{Q}(t-1) \\
\label{2sdepol2}
\hat{\rho}_{2s\sigma}(t) &= \frac{p}{3} \left[\sum_{i=1}^{3}
  \hat{\sigma}_i \hat{Q}_{\sigma}(t-1) \hat{\sigma}_i^{\dagger} \right
] + (1-p) \hat{Q}_{\sigma}(t-1),
\end{align}
\end{subequations}
where  $\hat{\sigma}_1 = \begin{bmatrix}0   &  1  \\
  1 & 0
\end{bmatrix}\otimes  \hat{{\mathbbm 1}}$, $\hat{\sigma}_2 = \begin{bmatrix}0   &  -i  \\ 
i &  ~~0 
\end{bmatrix}\otimes  \hat{{\mathbbm 1}}$, $\hat{\sigma}_3 = \begin{bmatrix}1   &  ~~0  \\ 
0 &  -1 
\end{bmatrix}\otimes \hat{{\mathbbm 1}}$ and $\hat{Q}(t-1)$ and
$\hat{Q}_{\sigma}(t-1)$ are given by Eqs.\,(\ref{q1}) and (\ref{q2}).
\par
\begin{figure}[tb]
\bc 
\subfigure[Alternate walk]{\includegraphics[width=8.6cm]{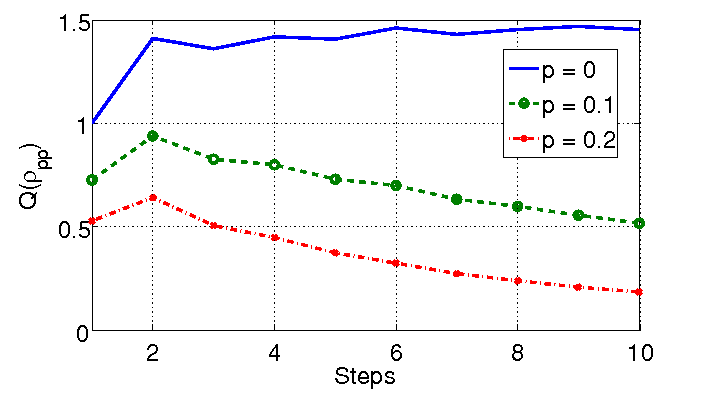}
\label{fig:13a}}
\subfigure[Pauli walk]{\includegraphics[width=8.6cm]{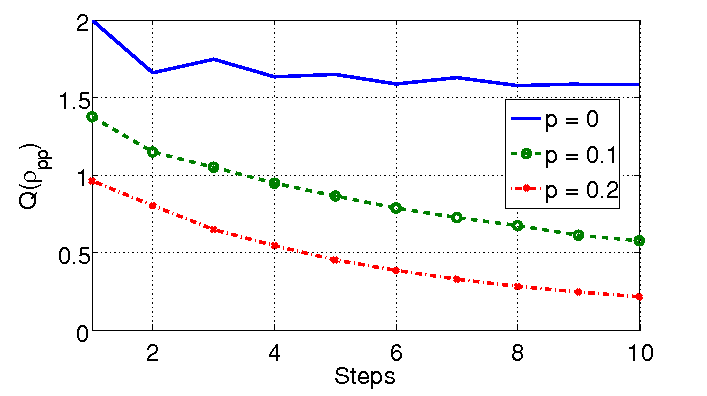}
\label{fig:13b}}
\ec
\caption{(Color online) Particle-position quantum correlations created by the two-state walks for different depolarizing noise levels.  
\label{fig:13}}
\end{figure}
\begin{figure}[ht]
\bc 
\subfigure[Alternate walk]{\includegraphics[width=8.6cm]{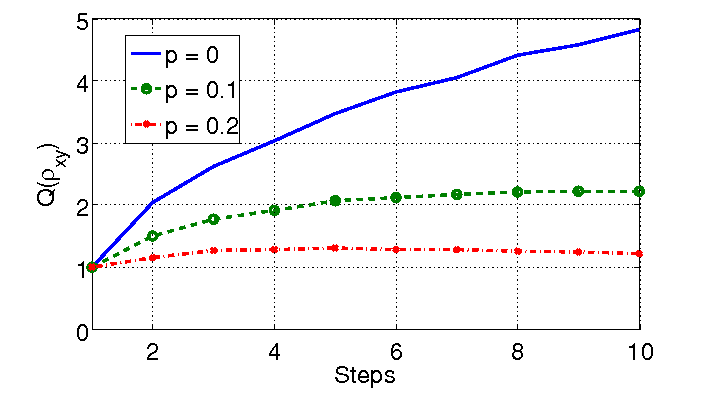}
\label{fig:14a}}
\subfigure[Pauli walk]{\includegraphics[width=8.6cm]{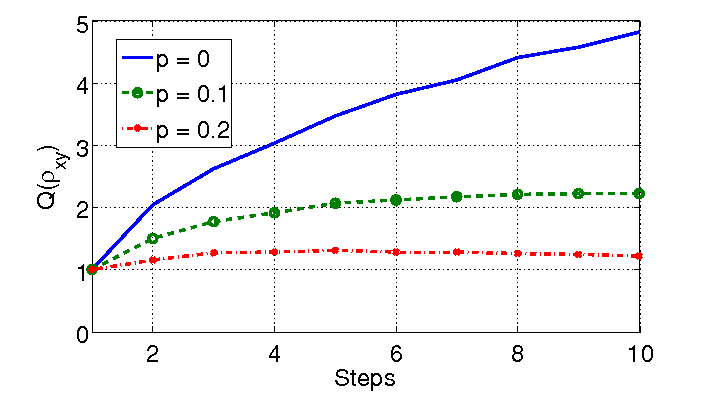}
\label{fig:14b}}
\ec
\caption{(Color online) Quantum correlations between the $x$ and $y$ spatial dimensions created by the two-state walks for different depolarizing noise levels.  \label{fig:14}}
\end{figure}

Similarly to the situation where we considered only state-flip noise after one complete step (see Figs.\,\ref{fig:8} and \ref{fig:9}) we find again that the quantum correlations for both walks behave nearly identical  and only differ slightly in strength compared to the case of state-flip noise alone (see Figs.\,\ref{fig:13} and \ref{fig:14}).  


\subsection{Robustness of two-state walk}
\label{robust}
\begin{figure}[ht]
\includegraphics[width=8.4cm]{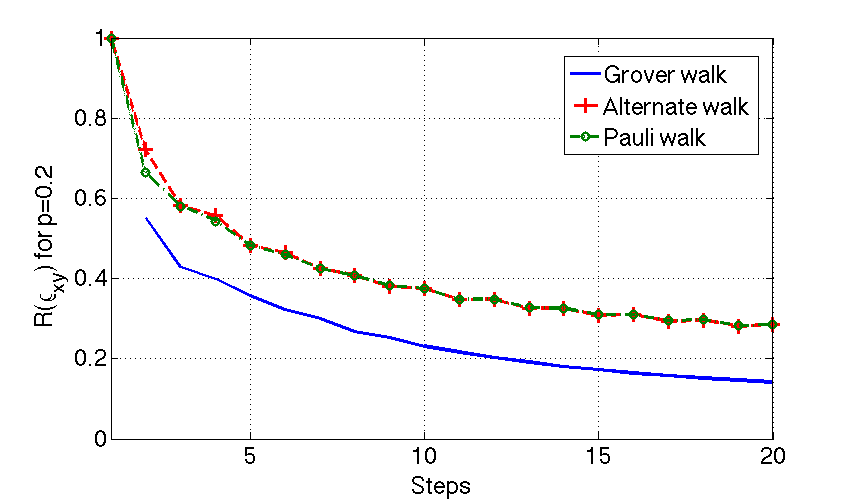}
\caption{(Color online) Relative decay of the quantum correlations $Q(\rho_{xy})$ in the presence of a state-flip noise channel with $p=0.2$. \label{fig:15}}
\end{figure}

\begin{figure}[ht]
\includegraphics[width=8.4cm]{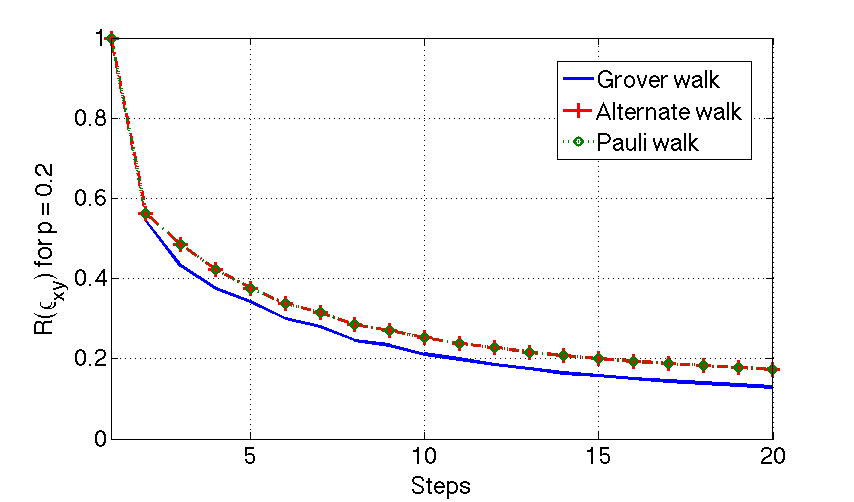}
\caption{(Color online) Relative decay of the quantum correlations $Q(\rho_{xy})$ in the presence of a depolarizing noise channel with $p=0.2$. \label{fig:16}}
\end{figure}

From the preceding sections we note that the $x-y$ spatial correlations, $Q(\rho_{xy})$, have a larger absolute value for the two-state walks compared to the Grover walk and that the presence of noise affects all schemes in a similar manner.  To quantify and better illustrate the effect the noise has we therefore calculate
\begin{equation}
	R(\rho_{xy}) = \frac{ Q(\rho_{xy})\quad\text {for noisy walk}}{Q(\rho_{xy}) \quad\text{for noiseless walk}},
\end{equation}
as a function of number of steps, which gives the rate of decrease in the quantum correlations. In Figs.\,\ref{fig:15} and \ref{fig:16} we show this quantity in the presence of state-flip or depolarizing noise, respectively, for a noise level of  $p=0.2$. One can clearly see that in both cases the two-state walks are more robust against the noise at any point during the evolution. Note that the Grover walk only produces correlations from step 2 on, which is the reason for its graph starting later.

\section{Breakdown of state-flip and phase-flip symmetries for
  four-state walks}
\label{sec:symm}

The quantum walk of a two-state particle in 1D is known to remain unaltered in the presence of unitary operations which equally effect each step of the evolution. This is due to the existence of symmetries\,\cite{CSB07, BSC08}, which can help to identify different variants of the same quantum walk protocol and which can be useful in designing experimental implementation. For example, in a recent 
scheme used to implement a one-dimensional quantum walk using atoms in an optical lattice\,\cite{Cha06}, the conditional shift operator also flipped the state of the atom with every shift in position
space. However, the existence of a bit-flip symmetry in the system allowed to implement the
walk without the need for compensation of these bit-flips. In this section we look at possible symmetries in the walks discussed above and show that the bit-flip and phase-flip symmetries, which are present in the evolution of the two-state particle are absent in the evolution of the four-state particle.

The density matrix for a two-state quantum walk in the presence of a noisy channel
will evolve through a linear combination of noisy operations on the state and
the unaffected state itself. As an example we illustrate the symmetry due to bit-flip operations in the alternate walk with bit-flip noise after evolution of one complete step. The density matrix in this case is given by
\bea
\hat{\rho}_{2s}(t) &=&  p  \Big [ \hat{\sigma}_1   \hat{S}_y \hat{S}_x \hat{\rho}_{2s}(t-1) \hat{S}_x^{\dagger} \hat{S}_y^{\dagger}  \hat{\sigma}^{\dagger}_1  \Big ] \nonumber  \\ 
& & +(1-p) \Big [ \hat{S}_y \hat{S}_x \hat{\rho}_{2s} (t-1) \hat{S}^{\dagger}_{x}\hat{S}^{\dagger}_{y} \Big ],
\eea
where $\hat{\sigma_1}  = \begin{bmatrix}0   &  1  \\
  1 & 0
\end{bmatrix}\otimes  \hat{{\mathbbm 1}}$.
When the noise level is $p=1$ this expression reduces to 
\bea
\label{eq:sysm}
\hat{\rho}_{2s}(t) &=&  \Big [ \hat{\sigma}_1   \hat{S}_y \hat{S}_x \hat{\rho}_{2s}(t-1) \hat{S}_x^{\dagger} \hat{S}_y^{\dagger}  \hat{\sigma}^{\dagger}_1  \Big ] \nonumber \\
 &=&\hat{S}^{\prime}_y \hat{S}_x \hat{\rho}_{2s}(t-1) \hat{S}_x^{\dagger} (\hat{S}_y^{\prime})^{\dagger}.  
\eea
where in the second line the bit-flip operation has been absorbed into the evolution operator $\hat{S}^{\prime}_{y}$. This replaces 
$|0\rangle \langle 0|$ and $|1\rangle \langle 1|$ in $\hat{S}_{y}$ by $|1\rangle \langle 0|$ and $|0\rangle \langle 1|$, respectively.  Similarly,  for a phase-flip, $\hat{\sigma_1}$ in Eq.\,(\ref{eq:sysm})  is replaced by $\hat{\sigma_3}  = \begin{bmatrix}1   &  0  \\
  0 & -1
\end{bmatrix}\otimes  \hat{{\mathbbm 1}}$ leading to
$|1\rangle \langle 1|$ in $\hat{S}_{y}$ being replaced by $-|1\rangle \langle 1|$  to construct $\hat{S}^{\prime}_{y}$.

An alternative way to look at this is to absorb the bit-flip or phase-flip operation into the coin operation. 
For a two-state walk using the Hadamard coin operation, the bit-flip operation after each steps corresponds to the coin operation taking the form, \bea \hat{H}^{\prime} =
\frac{1}{\sqrt{2}}\begin{bmatrix}
  1 & &-1 \\
  1 && ~~1
\end{bmatrix}
\eea
and the phase-flip after each steps  corresponds to the coin operation taking the form,
\bea
\hat{H}^{\prime \prime} = \frac{1}{\sqrt{2}}\begin{bmatrix}
~~1 & & 1 \\
-1 && 1 
\end{bmatrix}.
\eea
For a bit-flip or phase-flip of noise level $p=1$ the Hadamard coin operation $\hat{H}$ can therefore be recast into a noiseless ($p=0$) quantum walk evolution using $\hat{H}^{\prime}$ and $\hat{H}^{\prime \prime}$ as coin operation.  A noise level of $p=1$ then returns a probability distribution equivalent to the noiseless evolution and consequently the maximum bit-flip and phase-flip noise level for a two state walk corresponds to $p=0.5$. This is a symmetry within the alternate walk, which also holds for the Pauli walk.

A state-flip noise channel for the four-state walk, on the other hand, evolves the state into a linear combination of  all possible flips between the four basis states and an unchanged state for all values of $p$ except for $p=0$   (see Eq.\,\ref{4s23}).    That is, only when $p=0$, Eq.\,(\ref{4s23}) reduces to 
\begin{eqnarray}
\label{4s23a}
   \hat{\rho}_{4s}(t) =\hat{S}_4\hat{\rho}_{4s}(t-1)\hat{S}_4{^\dagger},
\end{eqnarray}
whereas for any non-zero $p$ including $p=1$, Eq.\,(\ref{4s23}) takes the form
\begin{eqnarray}
\label{4s23b}
   \hat{\rho}_{4s}(t) &=& \frac{1}{k} 
                         \left[\sum_{i =1}^k\hat{f}_i\hat{S}_4\hat{\rho}_{4s}(t-1)
                               \hat{S}_4{^\dagger} \hat{f}_i{^\dagger} \right],
\end{eqnarray}
Any attempt to absorb the noise operations $\hat{f}_i$ into the shift operator or the coin operation leads to $k$ different results, which have to be applied with probability $\frac{1}{k}$. Therefore, in contrast to the two-state evolution, a state flip noise level of $p=1$ does not result in a pure evolution equivalent to the situation for $p=0$. However, if the state-flip noise is restricted to one possible operation ($\hat f_i$), the density matrix is no longer a linear
combination of noisy operations and the unchanged state. Thus, for $p=1$ in
Eq.\,(\ref{4s23}) a single
noise operation can be absorbed into the GDO by changing the form of $G$
(see Eq.\,(\ref{grovercoin})). 

This absence of a useful symmetry for the
four-state quantum walk reduces the chances to find an
equivalent class of four-state quantum walk evolutions.  Furthermore, since the four-state quantum walk requires a specific
form of coin operation to implement the walk, any possible absorption will not result in a quantum walk in 2D, which is a significant difference to the two-state walks.

\section{Conclusion} 
\label{conc}

In this work we have studied the decoherence properties of three
different schemes that realize a quantum walk in two-dimensions, namely the Grover (four-state)
walk, the alternate walk and the Pauli walk. The noise for two-state
particle evolution was modeled using a bit-flip channel and
depolarizing channel. For the four-state evolution, different possible
state-flip channels were explored and we have shown a channel
with 3 cyclic flips between all the four states can be used as a very good approximation to the full situation. Similarly, we presented a possible model for the depolarizing channel of the four-state quantum walk. Using MID as a measure for the quantum correlations within the state, our studies have shown that the two-state quantum walk evolution is in general more robust against decoherence from state-flip and depolarizing noise channels. 

Following earlier studies on bit- and phase-flip symmetries in two-state quantum walks in 1D, we have shown that they also hold for two-state quantum walks in 2D, but break down
for four-state 2D quantum walks.

With the larger robustness against decoherence, the existence of symmetries which allow freedom of choice with respect to the initial state and the coin operation and the much easier experimental control, we conclude that two-state particles can be conveniently used to implement quantum walks in 2D compared to schemes using higher dimensional coins.  An other important point to be noted is the straightforward extendability of the both two-state schemes to higher dimensions
by successively carrying out the evolution in each dimension.

{\bf Acknowledgement:} We acknowledge support from Science Foundation
Ireland under Grant No. 10/IN.1/I2979. We would like to thank Carlo Di
Franco and Gianluca Giorgi for helpful discussions.


\end{document}